\documentclass[12pt]{article}
\usepackage{epsfig, array,multirow, bm, amsfonts, amssymb, amsmath}

\usepackage[greek,english]{babel}
\usepackage{graphicx}
  
\RequirePackage{color}

\usepackage{graphicx}
\RequirePackage{color}

\definecolor{MyDarkGreen}{rgb}{0.02,0.60,0.06}

\usepackage{soul}

\usepackage{teubner}

\def\qq{\hbox{\foreignlanguage{greek}{\coppa}}}
\def\qqq{{\hbox{\foreignlanguage{greek}{\footnotesize\coppa}}}}

\title{\bf Finite-size scaling above the upper critical dimension in Ising models with long-range interactions}
\author{ 
{\it E.J.\ Flores-Sola$^{\,1,2}$,} {\it B. Berche$^{\,2}$,} {\it R. Kenna$^{\,1}$} and {\it M. Weigel$^{\,1}$}\\~\\
$^1$ 
Applied Mathematics Research Centre,\\
Coventry University,\\
Coventry, CV1 5FB, England
{}\\~\\
$^2$ Statistical Physics Group, \\
Institut Jean Lamour, UMR CNRS 7198,\\
Universit{\'{e}} de Lorraine, \\
B.P. 70239, 54506 Vand\oe uvre l\`es Nancy Cedex, France
{}\\~\\}
\begin{document}
\maketitle


                      {\Large
                      \begin{abstract}
The correlation length plays a pivotal role in finite-size scaling and 
hyperscaling at continuous phase transitions. 
Below the upper critical dimension, where the correlation length is proportional to the system length, both finite-size scaling and hyperscaling take conventional forms. 
Above the upper critical dimension these forms break down and a new scaling scenario appears. 
Here we investigate this scaling behaviour by simulating
one-dimensional Ising ferromagnets with long-range interactions. 
We show that the correlation length scales as a non-trivial power of the linear system size and investigate the  scaling forms. 
For interactions of sufficiently long range, the disparity between the correlation length and the system length can be made arbitrarily large, while maintaining the new scaling scenarios.
We also investigate the behavior of the correlation function above the upper critical
dimension and the modifications imposed by the new scaling scenario onto the associated Fisher relation.
%
                        \end{abstract} }
%
  \thispagestyle{empty}
%
%
  \newpage
%
                  \pagenumbering{arabic}

\section{Introduction}
\setcounter{equation}{0}

It is frequently stated that hyperscaling fails in magnetic systems above the upper
critical dimension $d_c$.  The standard expression for the associated scaling
relation is $ \nu d = 2 - \alpha$. Clearly, if the exponents $\alpha$ and $\nu$
associated with the specific heat and correlation length, respectively, take on their
mean-field values, this  relation can only be valid for one value of $d$, and
not for all $d > d_c$. 
It has recently been shown, however, that above $d_c$,
hyperscaling can be restored \cite{BeKe12,ourCMP} by relaxing a previous
implicit assumption that the finite-size correlation length $\xi_L$ is bounded by the
linear system size $L$ \cite{BNPY}. If, instead, the correlation length scales
algebraically with $L$,
\begin{equation}
 \xi_L \sim L^{\qqq},
\label{koppa}
\end{equation}
hyperscaling holds in the form \cite{BeKe12,ourCMP}
\begin{equation}
 \frac{\nu d}{\qq} = 2 - \alpha.
\label{hyperscaling}
\end{equation} 
The exponent ${\qq}$ (``koppa'' \cite{BeKe12,ourCMP}) is $1$ if $d \le d_c$ so that 
standard hyperscaling is recovered there. It takes the value ${\qq} = 
d/d_c$ above the upper
critical dimension, so that the hyperscaling relation 
reduces to $ \nu d_c = 2 - \alpha$.
The combination $d/d_c$ has appeared explicitly or implicitly in earlier works with periodic boundaries \cite{Br82,Turks,JY05,KLY09,BWM12}.
In refs.~\cite{BeKe12,ourCMP}, however, it was shown that besides having 
physical significance, it is also universal.

In Ref.~\cite{ourEPL}, the decay of the correlation function above the upper critical
dimension was also revisited.  The generic form
\begin{equation}
 G(r) \sim \frac{1}{r^{d-2+\eta}}
\label{G1}
\end{equation}
is associated with  Fisher's scaling relation \cite{Fi64}
\begin{equation}
  \gamma = \nu (2-\eta),
	\label{G2}
\end{equation}
in which $\eta$ is the anomalous dimension.  It was shown
that, when measured on the finite-size system-length scale and properly taking dangerous
irrelevant variables into account above the upper critical dimension, the correct
scaling form is
\begin{equation}
 G(r) \sim \frac{1}{r^{d-2+\eta_Q}},
\label{G3}
\end{equation}
where $\eta_Q$ is related to $\eta$ through
\begin{equation}
  \eta_Q = \qq \eta + 2(1-\qq).
	\label{G4}
\end{equation}
Thus $\eta_Q$ reverts to $\eta$ when $\qq=1$ below the upper critical dimension.
This gives a new version of Fisher's scaling relation above $d_c$ as
\begin{equation}
  \qq \gamma  = \nu (2-\eta_Q).
	\label{G5}
\end{equation}

The main focus of this paper is on the new exponent ${\qq}$. Its non-triviality is in
disagreement with the traditional formulation of hyperscaling above the
upper critical dimension and with detailed finite-size scaling there, where it was
required that the correlation length be bounded by the system size~\cite{BNPY}.  Our
aim is to numerically verify Eqs.~(\ref{koppa}) and (\ref{G3}) for the Ising model
with interactions of long range.  We are particularly interested in such systems
because the long ranges can reduce the upper critical dimension from $d_c=4$ of the
short-range model to experimentally accessible values. Here, in particular, we
investigate an Ising system in $d=1$ dimension.

Besides direct tests of the relations (\ref{koppa}) and (\ref{G3}), we study
explicitly the finite-size scaling (FSS) of thermodynamic observables above the upper
critical dimension, where the exponent $\qq$ enters too.  Indeed, like hyperscaling,
the conventional form for FSS is transformed above $d_c$.  Let $P_L(t)$ represent an
observable $P$ measured for a system of linear extent $L$ at reduced 
temperature $t$.
If $P_\infty(t) \sim |t|^{-\rho}$, conventional FSS posits that $P_L(t) \sim
L^{{\rho}/{\nu}}$ inside the scaling window \cite{Ba83}.  Above the upper critical
dimension, however, this conventional form is replaced by
\begin{equation}
 P_L(t) \sim L^{\frac{\qqq \rho}{\nu}}.
 \label{QFSS}
\end{equation}
This is called $Q$-FSS to distinguish it from the conventional form
\cite{BeKe12,ourCMP,ourEPL}.  Thus $Q$-FSS contains information on the exponent $\qq$
and can be used to measure it.

The rest of this paper is organised as follows. In Section 2, we recall the physics
of spin models with long-range interactions.  Sections 3 and 4 discuss the
conventional and new pictures for scaling above the upper critical dimension. After
introducing the numerical techniques in Section 5, we present our simulation results
for the one-dimensional model in Section 6. Finally, Section 7 contains our
conclusions.

\section{Ising model with long-range interactions}
\setcounter{equation}{0}

We consider a ferromagnetic Ising model with Hamiltonian
\begin{equation}
 \mathcal{H}=- \sum_{i<j} J_{ij} s_i s_j+\sum_{i} H_i s_i,
\label{hmil}
\end{equation}
where $s_i$ represents the spin at site $i$ and $H$ denotes an external 
magnetic field. The coupling constant $J_{ij}$ is given by
\begin{equation}
 J_{ij}=\frac{J}{r_{ij}^{d+\sigma}},
 \label{eq:couplings}
\end{equation}
where $r_{ij}=|\vec{r}_i-\vec{r}_j|$ is the distance between spins $s_i$ and $s_j$.

The physics of this system was first systematically discussed by Fisher, Ma and
Nickel \cite{FMN72}. Their RG treatment identified a number of different regimes
in the model: for $0 < \sigma < \sigma_U = d/2$, the Gaussian fixed point is stable
and one expects mean-field behavior, i.e., the system is above its upper critical
dimension which is hence
\begin{equation}
 d_c = 2 \sigma.
\label{dc}
\end{equation}
Here, the critical exponents are found to be
\begin{eqnarray}
 \alpha & = & 0, \quad \beta = \frac{1}{2}, \quad \gamma = 1, \quad \delta = 3,
 \label{abcd} \\
 \nu  &= & 
\frac{1}{\sigma}, \quad {\mbox{and}} \quad \eta = 2 - \sigma.
 \label{ef}
\end{eqnarray}
For $d/2 < \sigma < \sigma_L = 2$ non-mean-field behavior is expected with critical
exponents changing continuously with $\sigma$. Finally, for $\sigma > 2$, the
behavior of the short-range model is recovered. Following this initial treatment,
there has been a protracted debate about the situation at the lower critical
$\sigma_L$, where short-range behavior is recovered. An excellent summary of this
development is given in the recent Ref.~\cite{AngeliniParisi}. Here, it suffices to
say that the lower critical range was later conjectured to be more precisely
$\sigma_L = 2-\eta_\mathrm{SR}$ \cite{Sak73,HN89}, where $\eta_\mathrm{SR}$ is the
correlation function exponent for the corresponding $d$-dimensional short-range
universality class. For $d=1$, in particular, this implies $\sigma_L = 1$, in
agreement with exact results for this specific case \cite{FS82}. Recent discussions
have focused on the location of and behavior at the lower critical $\sigma_L$
\cite{AngeliniParisi,Picco2012,BlPi13,BrezinParisi}. Here, however, we are interested
in the classical regime to show that long-held assumptions regarding the correlation
length are incorrect.  We provide evidence that the correct scaling picture is that
provided by $Q$-FSS.

Throughout this study, we restrict ourselves to a one-dimensional chain with periodic
boundary conditions, i.e., $d=1$. The relevant distance between spins is then more
precisely given by
\begin{equation}
r_{ij}^{\shortmid}=\min(|i-j|,L-|i-j|).
        \label{Eq:spin_dist_1d_pow_line}
\end{equation}
As a result of the long-range nature of interactions and the periodic boundary, the
coupling of each spin to an infinite number of replicas of each partner spin at
larger and larger distances must be taken into account, leading to renormalized
couplings
\begin{equation}
  \tilde{J}_{ij}  = \sum_{n=-\infty}^{\infty} \frac{1}{|r_{ij}^{\shortmid}+Ln|^{1+\sigma}}
  = \frac{1}{|L|^{1+\sigma}} \left[\zeta\left(1+\sigma,\frac{r_{ij}^{\shortmid}}{L}\right) + 
                                        \zeta\left(1+\sigma,1-\frac{r_{ij}^{\shortmid}}{L}\right) \right],
\label{eq:resummed}
\end{equation}
with the Hurwitz Zeta function \cite{AS65}
\begin{equation}
        \zeta\left(s,q\right) :=  \sum_{k=0}^{\infty} \frac{1}{(k+q)^{s}}.
\end{equation}

\section{Scaling above the upper critical dimension: old picture}
\setcounter{equation}{0}

The traditional hyperscaling relation originates in Widom's universal scaling
hypothesis that the 
thermodynamic functions depend homogeneously on the reduced temperature $t = 1-T/T_c$
(where $T_c$ is the critical value of $T$) and magnetic field $h$ (which is zero at
the critical point) \cite{history}.  The extension of this idea to finite-size
systems provides a grounding for FSS theory \cite{Ba83}.  The homogeneity assumptions
for the free energy, correlation length and correlation function are
\begin{eqnarray}
f_L(t,h,u) & = &b^{-d}f_{L/b}(tb^{y_t},h b^{y_h}, u b^{y_u}),
\label{fefss}\\
\xi_L(t,h,u) & = &b \xi_{L/b}(tb^{y_t},h b^{y_h}, u b^{y_u}),
\label{fefss2} \\
G_L(t,u,r) & = & b^{-2X_\phi} G_{L/b}(tb^{y_t}, u b^{y_u}, rb^{-1}),
\label{fefss3}
\end{eqnarray}
respectively.  Here $u$ represents a parameter in the Hamiltonian which, in the case
of $\phi^4$ theory, is the bare quartic coupling.  Below $d_c$, non-trivial critical
behavior is defined by the Wilson-Fisher fixed point and irrelevant scaling fields
lead to Wegner corrections \cite{Wegner72}.  In that case, setting the rescaling
factor $b=L$ and $h=0$ in Eq.~(\ref{fefss2}) allows one to identify $\xi_\infty (t,0)
\sim t^{-\nu}$ with $\nu = 1/y_t$ on the one hand, and $\xi_L(0,0) \sim L$ on the
other.  Eq.~(\ref{fefss}) then gives the finite-size free energy to be a function of
$L/\xi_\infty$, so that this ratio controls finite-size scaling (FSS) below $d_c$.
Twice differentiating the scaling form $f_\infty (t,0) \sim t^{\nu d}$ then leads to
the standard hyperscaling relation $\nu d = 2 - \alpha$. 

Above the upper critical dimension, the critical behaviour is determined by the
Gaussian fixed point. The scaling dimensions for the long-range Ising model there are
\cite{FMN72}
\begin{equation}
 y_t=\sigma, \quad y_h=\frac{d+\sigma}{2} \quad {\mbox{and}} \quad
y_u=2\sigma-d.
\label{sdim}
\end{equation}
Above $d_c=2\sigma$, $u$ becomes irrelevant.  However, it can also be dangerous and
therefore cannot simply be set to zero \cite{FiHa83}.  This means that the above
forms of FSS and hyperscaling both break down above $d_c$.  Using homogeneity, we
write Eq.~(\ref{fefss}) as
\begin{equation}
f_L(t,h,u)  = b^{-d}\tilde{f}_{L/b}(tb^{y_t^*},h b^{y_h^*}).
\label{Emilio1}
\end{equation}
If
\begin{equation}
 y^*_t=y_t-\frac{y_u}{2}=\frac{d}{2}, \quad
{\mbox{and}} \quad
y^*_h=y_h-\frac{y_u}{4}=\frac{3d}{4},
\label{starred}
\end{equation}
this form recovers the correct, Gaussian scaling behaviour (\ref{abcd}) for the
thermodynamic functions in the thermodynamic limit.
One notes that Eqs.(\ref{starred}) are $\sigma$-independent, unlike the scaling dimensions in Eqs.(\ref{sdim}).

It is well established that the correlation length critical exponent value for the
long-range model above $\sigma_U = d/2$ is $\nu = 1/\sigma$.  This coincides with the
expectation $\nu = 1/y_t$, without recourse to dangerous irrelevant variables.  For
this reason, $u$ was believed not to be dangerous for the correlation length and
hence could safely be set to zero in Eq.~(\ref{fefss2}). That equation would then
become
\begin{equation}
\xi_L(t,h,0)  =  b  \xi_{L/b}(tb^{y_t},h b^{y_h}, 0),
\label{fefss22}
\end{equation}
Setting $t=h=0$ and $b=L$, one obtains $\xi_L \sim L$, in accordance with the belief
that the correlation length cannot exceed the length of the system \cite{BNPY}.

FSS for the thermodynamic functions is now controlled by the first argument on the
right hand side of Eq.~(\ref{Emilio1}), and not by the combinations appearing in
Eq.~(\ref{fefss22}).  Observing that the argument is a function of the ratio
$t^{-1/y_t^*}/L$, Binder introduced the notion of the thermodynamic length, defined
as $\ell_\infty(t) \sim t^{-1/y_t^*}$ \cite{Bi85}.  This picture of scaling above the
upper critical dimension involves a number of length scales.  Besides the finite-size
system length $L$, one has the correlation length $\xi_\infty(t) \sim t^{-\nu}$ with
$\nu = 1/y_t$, the thermodynamic length $\ell_\infty(t) \sim t^{-1/y_t^*}$ and their
finite-size counterparts \cite{BDT}. 

Similarly, if $u$ is not dangerous for the correlation function, one may set $u=0$ in
Eq.~(\ref{fefss3}) at criticality and set $b=r$ to obtain $G_L(0,0,r) \sim
r^{-2X_\phi} G_{L/r}(0,0,1)$. Writing $G_{L/r}(0,0,1)$ as $g(r/L)$ and taking the
thermodynamic limit, one finds the asymptotic behavior $G_\infty(0,0,r) \sim
r^{-2X_\Phi}$. Comparing with the generic form (\ref{G1}), one has
\begin{equation}
 X_\phi = \frac{d-2+\eta}{2} = \frac{d-\sigma}{2},
\end{equation}
so that Eq.~(\ref{fefss3}) is
\begin{equation}
 G_L(0,0,r) = \frac{1}{r^{d-2+\eta}} g\left({\frac{r}{L}}\right),
\label{G6}
\end{equation}
with $\eta = 2-\sigma$.
The correlation function is related to the susceptibility through the
fluctuation-dissipation theorem, and this relationship delivers Fisher's scaling
relation (\ref{G2}).

\section{Scaling above the upper critical dimension: new picture}
\setcounter{equation}{0}

This established picture was challenged in Refs.~\cite{BeKe12,ourCMP,ourEPL}, where
the restriction that the correlation length be bounded by the system size \cite{BNPY}
was relaxed.  The homogeneity argument then allows one to write Eq.~(\ref{fefss2}) as
\begin{equation}
\xi_L(t,h,u)  =  b^{\qqq} \tilde{\xi}_{L/b}(tb^{y_t^*},h b^{y_h^*}).
\label{Emilio2}
\end{equation}
Taking the infinite-volume limit, one obtains $\xi_\infty(t) \sim t^{-\qqq/y_t^*}$, 
so that $\nu = \qqq/y_t^*$. 
For this to agree with the established result $\nu = 1/y_t$, one requires that
\begin{equation}
 \qq = \frac{y_t^*}{y_t} = \frac{d}{2\sigma} = \frac{d}{d_c}.
 \label{Emilio4}
\end{equation}
Keeping $L$ finite in Eq.~(\ref{Emilio2}) and setting $t=h=0$, one obtains
Eq.~(\ref{koppa}).  In this picture, dubbed $Q$-scaling in
Refs.~\cite{BeKe12,ourCMP,ourEPL}, the notion of an extra thermodynamic length is
abandoned as unnecessary but so too is the notion that the finite-size correlation
length be bounded by the system length.  One of our objectives here is to verify this
for the Ising model in $d=1$ dimension with long-range interactions.

$Q$-scaling also delivers hyperscaling above the upper critical dimension
\cite{BeKe12,ourCMP}.  Differentiating Eq.~(\ref{Emilio1}) twice with respect to $t$,
setting $h=0$ and taking the limit $L \rightarrow \infty$, one obtains $c_\infty \sim
t^{-(2-d/y_t^*)}$ for the specific heat. Identifying the exponent as $-\alpha$,
Eq.~(\ref{Emilio4}) gives Eq.~(\ref{hyperscaling}), as proposed for the short-range
model in Refs.~\cite{BeKe12,ourCMP}.

Moreover, inserting $b=L$ in Eqs.~(\ref{Emilio1}) and (\ref{Emilio2}), one sees that
finite-size scaling is governed by the ratio $tL^{y_t^*} = L^{\qqq}/t^{-\nu} =
\xi_L(0)/\xi_\infty(t)$, without recourse to a new length scale $\ell$.  Below the
upper critical dimension, where $\qq=1$, this ratio becomes $L/\xi_\infty(t)$ and the
$Q$-version of finite-size scaling ($Q$-FSS) reverts to ordinary FSS.  The $Q$-FSS
forms for the magnetisation and susceptibility are
\begin{eqnarray}
 m_L & \sim & L^{-\frac{\qqq \beta}{\nu}} = L^{- \frac{\qqq \sigma}{2}} = L^{-\frac{d}{4}} ,
\label{mL}
\\
 \chi_L & \sim & L^{\frac{\qqq \gamma}{\nu}} = L^{\qqq \sigma} = L^{\frac{d}{2}} .
\label{chiL}
\end{eqnarray}
Finally, from Eq.~(\ref{Emilio1}), one expects a given thermodynamic function (e.g.,
the susceptibility) to have a finite-size peak when $t=t_L$ where $t_LL^{y_t^*} \sim
1$. This means that the peak position scales as 
\begin{equation}
  t_L  \sim L^{-\lambda}
\quad {\mbox{where}} \quad
\lambda = y_t^* = \frac{\qq}{\nu} = \frac{d}{2}.
\label{shift}
\end{equation}

Having now seen that dangerous irrelevant variables are, in fact, important for the
correlation length, we revisit the correlation function too.  Following
Ref.~\cite{ourEPL} we write Eq.~(\ref{fefss3}) as $G_L(t,u,r) = b^{-2X_\phi + v_1
  y_u} \tilde{G}_{L/b}(tL^{y_t^*}, rb^{-1+v_2y_u})$.  Setting $v_2=0$ to render
$rb^{-1}$ dimensionless and setting $v_1=-1/2$ after Ref.~\cite{ourEPL}, one obtains
\begin{equation}
 G_L(0,0,r) \sim \frac{1}{r^{d-2+\eta_Q}}g\left({\frac{r}{L}}\right),
\label{G11}
\end{equation}
where $\eta_Q = 2 - \sigma + y_u/2 = 2-d/2$ as in Eq.~(\ref{G4}).  One can check this
formula by differentiating the free energy (\ref{Emilio1}) with respect to two local
fields. This is the route to the scaling of the correlation function used in
Ref.~\cite{LuBl97b}.  Finally, applying the fluctuation dissipation theorem to
Eq.~(\ref{G11}) gives the Fisher-type relation (\ref{G5}).

Our objective in the remainder of this paper is to test Eqs.~(\ref{koppa}),
(\ref{mL}), (\ref{chiL}), (\ref{shift}) and (\ref{G11}) from a numerical simulation
of the $d=1$ Ising model with long-range interactions, tuning the interaction range
to the regime $\sigma < \sigma_U = 1/2$ corresponding to a system above its upper
critical dimension.  We are especially interested in Eq.~(\ref{koppa}) and will show
that the correlation length can, indeed, be arbitrarily larger than the system
length, as predicted by $Q$ theory. Similarly, we will investigate the correlation
function and show that it follows Eq.~(\ref{G11}) rather than the standard mean-field
prediction (\ref{G1}).

\section{Cluster-update Monte Carlo simulations}
\setcounter{equation}{0}

The long-range nature of the interactions (\ref{eq:couplings}) appears to require the
calculation of $N = L^d$ energy terms for updating the state of a single spin, such
that a full system update becomes an O($N^2$) operation. Furthermore, we are
interested in the critical behavior of the models considered, so we expect additional
critical slowing down to affect any Markov chain Monte Carlo simulation
\cite{BL09}. Cluster updates such as the Swendsen-Wang algorithm \cite{SW87}, based
on the Fortuin-Kasteleyn representation of the Ising model, are known to dramatically
reduce critical slowing down for models with short-range interactions. As was
demonstrated by Luijten \cite{Luijtenthesis}, however, similar algorithms can be even
more efficient for the long-range interactions discussed here, as additional to a
reduction of autocorrelation times, a full update of the configuration can be
performed in O($N\log N$) instead of the naive O($N^2$) operations.  To achieve this,
instead of considering, in turn, addition of the $N-1$ neighbors of a given spin to a
growing cluster, the algorithm samples directly from the {\em cumulative\/}
distribution of such events, deciding at which distance the next spin will be
successfully added \cite{LB95}.  Due to the tremendous speed-up achieved, this
approach has been used for virtually all subsequent studies of the long-range Ising
model, including the very recent studies mentioned above
\cite{AngeliniParisi,Picco2012}.

Here, we use an alternative technique suggested by Fukui and Todo \cite{FT09}.  It is
based on a slight generalization of the established Fortuin-Kasteleyn representation
\cite{FK72}.  In this classical formulation, binary bond variables $n_\ell \in \{0,
1\}$ are introduced, such that $n_\ell = 1$ (bond active) with probability $p_\ell =
1-e^{-2\beta J_\ell}$ if the spins on the two ends of the bond point in the same
direction and $n_\ell = 0$ (bond inactive or deleted) otherwise.  It was noted by
Luijten and Bl\"ote \cite{LB95} that it is permissible, alternatively, to choose
$n_\ell = 0$ or $1$ independent of the spin orientations according to $p_\ell$ first,
but only connect spins for parallel spin pairs afterwards.  We can generalize the
$n_\ell$ to have arbitrary non-negative integer values with the convention that any
$n_\ell \ge 1$ simply corresponds to an active bond and ensuring that the probability
for the event $n_\ell \ge 1$ is the same as that of $n_\ell = 1$ in the binary model,
i.e.,
\begin{equation}
\sum_{n=1}^\infty f(n_\ell) = p_\ell,
\label{eq:normalization_condition}
\end{equation}
where $f(n)$ is the probability distribution of $n_\ell$ for parallel spins. Choosing
$f(n)$ to be Poissonian,
\[
f(n) = \frac{e^{-\lambda}\lambda^n}{n!},
\]
the condition (\ref{eq:normalization_condition}) implies that $\lambda_\ell = 2\beta
J_\ell$. The beauty of this generalization is that, as the sum of Poissonian
variables is Poissonian as well, it suffices to draw the sum $n_\mathrm{tot} =
\sum_\ell n_\ell$ from a Poisson distribution with mean $\lambda_\mathrm{tot} =
\sum_\ell \lambda_\ell$ and then distribute $n_\ell$ of these events to each bond
with a weight proportional to $\lambda_\ell$. This can be achieved in constant time
using Walker's method of alias \cite{Knu97}. As a result, the bond configuration
can be determined with O($N$) computational effort. Here, we use a tree-based
union-find data structure to perform successive cluster aggregation, which features
(almost) constant-time effort \cite{NZ01}. However, a single-cluster 
variant
can also be implemented with strictly O($N$) run-time scaling and is hence found to
be asymptotically more efficient than Luijten's approach.

For the simulations close to criticality, we determine integrated autocorrelation
times to ensure equilibration and sufficient independence of successive samples
\cite{Jan96}. Our simulations indicate a dramatic reduction of autocorrelation
times and also the dynamical critical exponents through the use of the cluster
updates. A detailed study of the dynamical critical behavior, in particular in the
region below the critical range parameter $\sigma_U = 1/2$, will be presented
elsewhere. Here, we investigate the scaling of the magnetization,
\begin{equation}
 m_L(t) = \frac{1}{L}
 \langle{|M|}\rangle,
\end{equation}
where $M = \sum_i s_i$ as well as the associated susceptibility,
\begin{equation}
 \chi_L(t) = \frac{1}{L}\left({
\langle{M^2}\rangle - \langle{|M|}\rangle^2
}\right).
\end{equation}
Their scaling gives access to the magnetic sector of the model. From the cluster
dynamics, improved estimators for these quantities are available as discussed in
Ref.~\cite{AngeliniParisi}.

To access the energetic sector and directly investigate the relevant length scale, we
extracted the second-moment correlation length. This can be determined from the
spin-spin correlation function
\begin{equation}
G(t, h, r) = \frac{1}{N} \sum_i \left(\langle s_i s_{i+r}\rangle - \langle s_i\rangle \langle
  s_{i+r}\rangle \right),
\label{eq:corr_fn}
\end{equation}
where periodic boundaries have been assumed. For long-range interactions, one expects
a modified Ornstein-Zernicke form of the propagator \cite{Suz73},
\[
\hat{G}(k) \sim \frac{1}{m^2+k^2+k^\sigma},
\]
where for $\sigma < 2$ the $k^\sigma$ is the dominant long wavelength
contribution. Hence, the correlation length can be estimated from
\cite{BCF00}
\begin{equation}
  \xi_L(t,h) = \frac{1}{2\sin(k_{\rm{min}}/2)}
  \left[{
      \frac{\hat{G}(0)}{\hat{G}(k_{\rm{min}})}-1
    }\right]^\frac{1}{\sigma}.
  \label{xinum}
\end{equation}
Here, $k_{\rm{min}}=2\pi/L$ is chosen to be the smallest wave vector for the periodic
lattice. The correlation function (\ref{eq:corr_fn}) itself is estimated here from
sampling a few long wave-length modes in reciprocal space and a final
back-transformation to real space.

\begin{figure}[t]
\begin{center}
\includegraphics[width=0.9\columnwidth, angle=0]{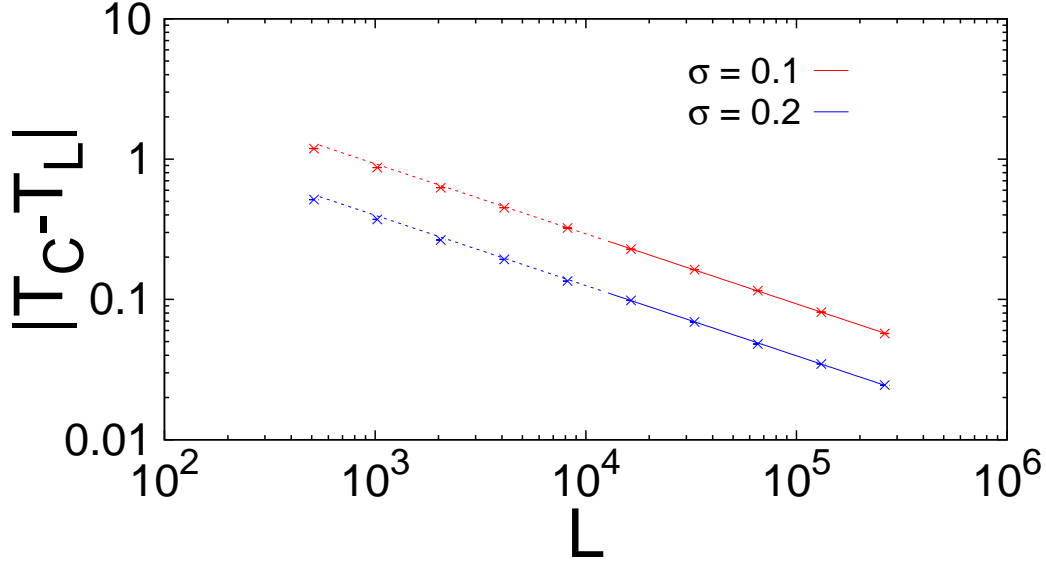}
\caption{Determination of the shift exponent $\lambda$ through the scaling of 
the
  susceptibility maxima $T_L$ for $\sigma = 0.1$ (top set of data, red online) and
  $\sigma = 0.2$ (lower set of data, blue online). The lines show fits of the
  functional form \eqref{eq:shiftform} to the data, where the range $L\ge 2^{14}$
  included in the fits is indicated by the solid part of the lines. The slopes
  estimate $-\qq/\nu$ as $-0.499 \pm 0.001$ and $-0.501 \pm 0.001$,
  respectively.}\label{figt}
\end{center}
\end{figure}

We note in passing, that naive measurements of the system energy and derived
quantities are O($N^2$) operations and hence costly. An O($N$) approach based on the
generalized Fortuin-Kasteleyn representation has been suggested in Ref.~\cite{FT09}.

\section{Simulation results}
\setcounter{equation}{0}

\begin{figure}[t]
\begin{center}
\includegraphics[width=0.9\columnwidth,angle=0]{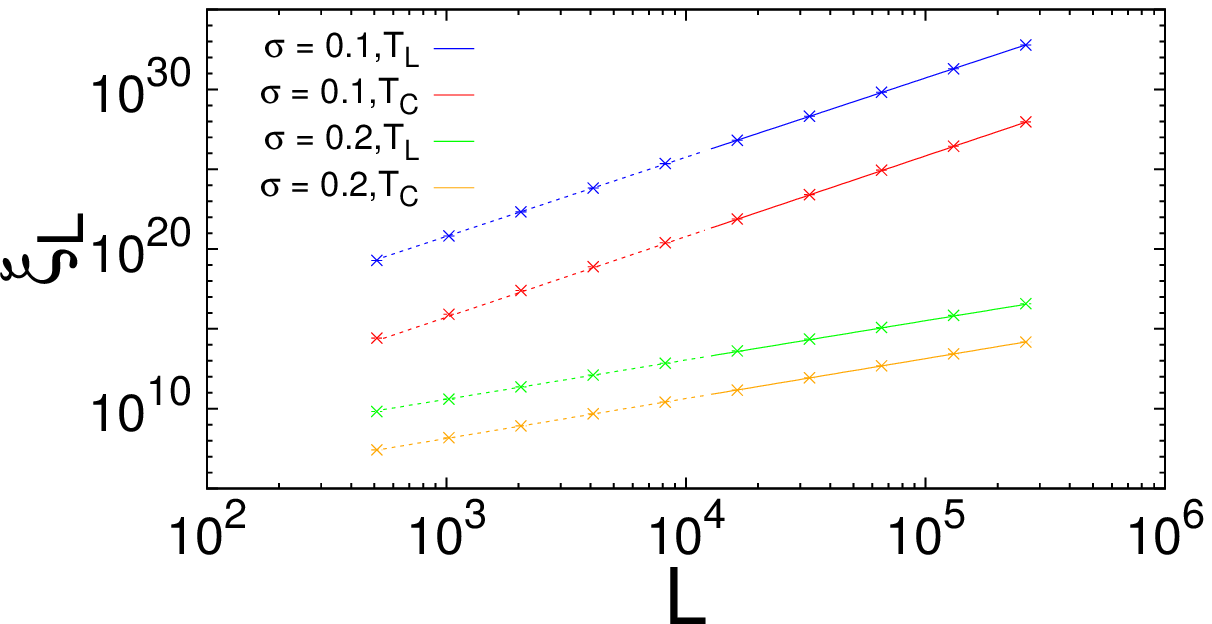}
\caption{Determination of the new exponent $\qq$ from the finite-size scaling
  of the correlation length (\ref{xinum}) for 
  $\sigma = 0.1$ at the pseudocritical point (top set of data, blue online),
  $\sigma = 0.1$ at the critical point (second from top, red online),
  $\sigma = 0.2$ at the pseudocritical point (third from top, green online),
  and $\sigma = 0.2$ at the critical point (lowest set of data, orange online).
  The slopes estimate $\qq$ as  
  $4.96 \pm 0.03$ and  
  $5.03 \pm 0.02$ for $\sigma = 0.1$, as well as 
  $2.50 \pm 0.02$ and
  $2.49 \pm 0.02$ for $\sigma = 0.2$, respectively.
}\label{figxi}
\end{center}
\end{figure}

\begin{table}[tb!]
\centering
 \begin{tabular}{|l|l|c|c|c|c|c|}  
 \hline
\multicolumn{2}{|c|}{}                                                     & $t_L$                                    & $\xi_L$                               &  $m_L$                                & $\chi_L$                           & $G_L(L/2)$                  \\ 
\multicolumn{2}{ |c| }{FSS}                                                & $L^{-\frac{1}{\nu}} = L^{-\sigma}$       & $L$                                   & $L^{-\frac{\beta }{ \nu}} = L^{-\frac{\sigma}{2}}$         & $L^{\frac{\gamma}{ \nu}} = L^{\sigma}$     & $L^{-(d-2+\eta)} = L^{\sigma - 1}$ \\
\multicolumn{2}{ |c| }{$Q$-FSS}                                            & $L^{-\frac{\qqq}{\nu}} =L^{-\frac{1}{2}}$& $L^{\qqq}=L^{\frac{1}{2\sigma}}$      & $L^{-\frac{\qqq\beta}{\nu}}=L^{-\frac{1}{4}}$              & $L^{\frac{\qqq\gamma}{\nu}}=L^{\frac{1}{2}}$& $L^{-(d-2+\eta_Q)}=L^{-\frac{1}{2}}$\\
\hline 
\parbox[t]{1mm}{\multirow{3}{*}{\rotatebox[origin=l]{90}{$\sigma = 0.1$}}} &     
                                     & -0.499 $\pm$ 0.001                     & 
 
                                     &                                       &   
                                 &       \\
                                                                     &$T_c$&     
                                     & 5.03 $\pm$ 0.02                     & 
-0.248  $\pm$ 0.001                    & 0.503 $\pm$ 0.002  &           -0.498 
$\pm$ 0.003 \\ 
                                                                     & $T_L$     
                                     &                                       & 
4.96 $\pm$ 0.03                     & -0.248 $\pm$ 0.001  & 0.504  $\pm$ 0.002 
                & -0.496 $\pm$ 0.001           \\ 
\hline                   
\parbox[t]{1mm}{\multirow{3}{*}{\rotatebox[origin=l]{90}{$\sigma = 0.2$}}} &     
                                     & -0.501 $\pm$ 0.001                     & 
 
                                     &                    &                      
              &  \\ 
                                                                    & $T_c$&     
                                     & 2.49 $\pm$ 0.02                     & 
-0.249 $\pm$ 0.001                     & 0.503 $\pm$ 0.002  & -0.490 $\pm$ 
0.002 
       \\ 
                                                                    & $T_L$&     
                                     &  2.50 $\pm$ 0.02                    & 
-0.246 $\pm$ 0.001                     & 0.508 $\pm$ 0.002  & -0.491 $\pm$ 
0.004 
            \\ 
\hline
	  \end{tabular}
    \caption{The conventional FSS and $Q$-FSS predictions {(top row)} as 
well as our numerical
      determination of various exponents for the 1D Ising model with long-range 
      interactions with $\sigma = 0.1$ {(second row)} and $0.2$ 
{(third row)}. {In each case, numerical estimates are given for both 
the critical point and the pseudocritical point}. The measured values fully 
support $Q$-FSS.}
    \label{tab1}
\end{table}

We performed simulations of the model (\ref{hmil}) with re-summed couplings
(\ref{eq:resummed}) using chains of lengths $L = 2^{9}$ to $L = 2^{18}$, initially
using a wide range of temperatures. The considered interaction ranges were
$\sigma = 0.1$ and $0.2$, deep in the mean-field region. Equilibration times and
measurement frequencies were set according to an analysis of integrated
autocorrelation times \cite{BL09}, resulting in up to $10^5$ Monte Carlo steps for
thermalisation, followed by $2 \times 10^5$ measurements.

Studying the magnetic susceptibility according to Eq.~(\ref{chiL}), we determined
pseudocritical temperatures as the location of the maxima of $\chi_L$,
\begin{equation}
  T_L = \operatorname*{arg\,max}_{T}\chi_L(T).
\end{equation}
Histogram reweighting \cite{BL09} was used to track the locations of these maxima,
iterating the simulation temperatures up to three times to ensure the absence of
reweighting bias. To determine the location of the critical point, we fitted the
shift equation
\begin{equation}
   T_L=T_c + A_t L ^{-\lambda}
   \label{eq:shiftform}
\end{equation}
to the data, initially using $T_c$, $A_t$ and the shift exponent $\lambda$ as free
parameters. To accommodate the presence of scaling corrections, which we do not
include here explicitly, we successively removed system sizes from the small-$L$ end
until satisfactory fit qualities were achieved.  This resulted in system sizes $L \ge
2^{14}$ being included in the fit. The data for the pseudocritical points together
with these fits are shown in Fig.~\ref{figt}.  The resulting parameters are $T_c=
21.006 \pm 0.001 $ with $\lambda=0.479 \pm 0.006$ for $\sigma=0.1$, and $T_c = 10.841
\pm 0.008$ with $\lambda=0.519 \pm 0.010$ for $\sigma=0.2$, respectively. As 
is clearly seen, the estimates of the shift exponent $\lambda$
are fully compatible with the $Q$-FSS prediction (\ref{shift}), $\lambda = 1/2$, but
not the conventional FSS result of $\lambda = 1/\nu = \sigma$.  Having established
confidence in the value of $\lambda$, in a second step we repeated the fits of the
form (\ref{eq:shiftform}) while fixing $\lambda = 1/2$, resulting in the more precise
estimates $T_c = 21.0013 \pm 0.0003$ ($\sigma = 0.1$) and $T_c = 10.8421 \pm 
0.0002$
($\sigma = 0.2$). We employ these values for the scaling analysis at
criticality. The
results are summarized in Table~\ref{tab1}. We note that our 
estimates of the transition temperatures are in
complete agreement with those reported in Ref.~\cite{Luijtenthesis}.

\begin{figure}[t]
\begin{center}
\includegraphics[width=0.9\columnwidth, angle=0]{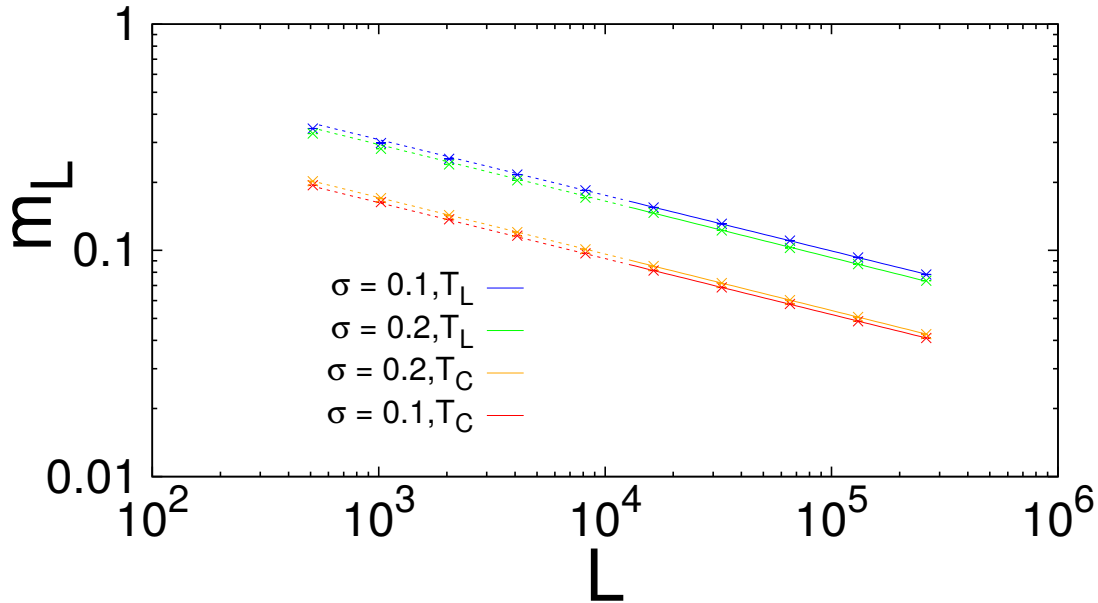}
\caption{Finite-size scaling of the magnetisation for 
$\sigma = 0.1$ at the pseudocritical point (top set of data, blue online),
$\sigma = 0.2$ at the pseudocritical point (second from top, green online),
$\sigma = 0.2$ at the critical point (third from top, orange online),
$\sigma = 0.1$ at the critical point (lowest set of data, red online).
The estimates of the critical exponent combination $-\qq\beta/\nu$ from fits of the
functional form (\ref{mL}) to the data are 
$-0.248 \pm 0.001$,
$-0.246 \pm 0.001$,  
$-0.249 \pm 0.001$ 
and
$-0.248 \pm 0.001$,
respectively.
}\label{figm}
\end{center}
\end{figure}

\begin{figure}[t]
\begin{center}
\includegraphics[width=0.9\columnwidth, angle=0]{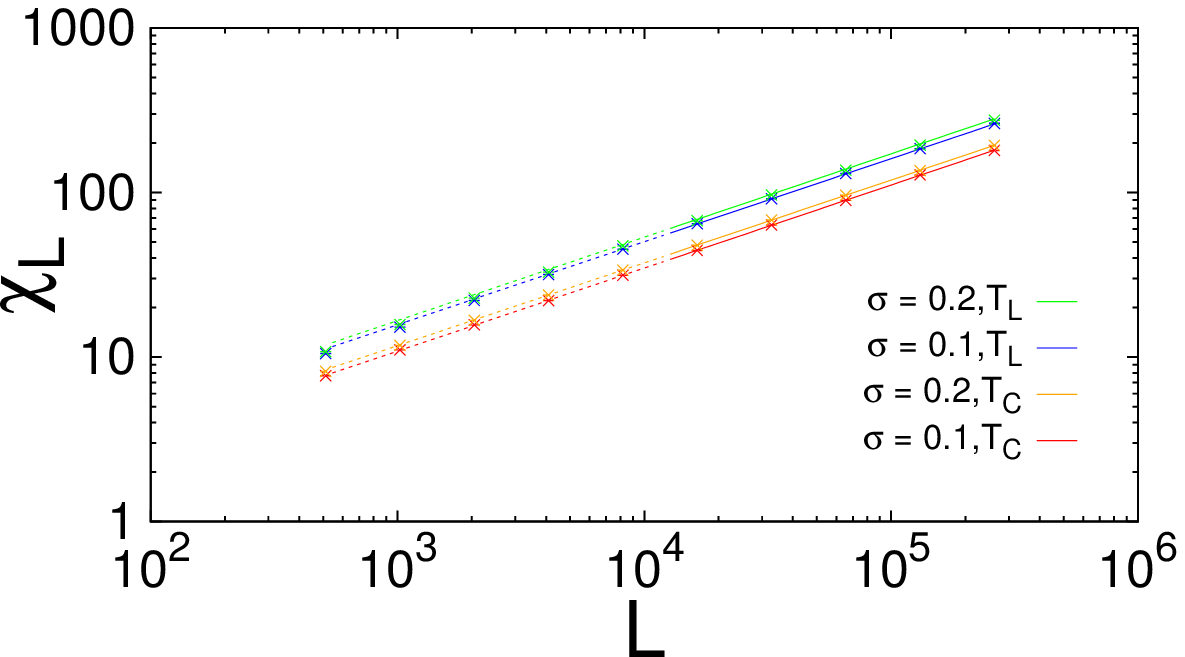}
\caption{ Finite-size scaling of the susceptibility in the 1d long-range Ising
  model. The data are for $\sigma = 0.2$ at the pseudocritical point (top set of
  data, green online), $\sigma = 0.1$ at the pseudocritical point (second from top,
  blue online), $\sigma = 0.2$ at the critical point (third from top, orange online),
  and $\sigma = 0.1$ at the critical point (lowest set of data, red online).  Fits of
  the functional form \eqref{chiL} to the data result in exponent estimates for
  $\qq\gamma/\nu$ of $0.508 \pm 0.002$, $0.504 \pm 0.002$, $0.503 \pm 0.002$ and
  $0.503 \pm 0.002$, respectively.  }\label{figchi}
\end{center}
\end{figure}

We now turn to the scaling of the correlation length as estimated from the
second-moment form \eqref{xinum}.  Figure~\ref{figxi} summarizes our results for the
correlation length for $\sigma = 0.1$ and $\sigma = 0.2$. For the critical point and
the data at the pseudocritical points $T_L$ defined from the susceptibility, we fit
the functional form \eqref{koppa} to the data for the size range $L\ge 2^{14}$. The
fits deliver $\qq = 5.03 \pm 0.02$ ($\sigma = 0.1$) and $\qq = 2.49 \pm 0.002$
($\sigma = 0.2$) for $t = 0$ and $\qq = 4.96 \pm 0.03$ ($\sigma = 0.1$) and $\qq =
2.50 \pm 0.02$ ($\sigma = 0.2$) for $t = t_L$.  This is in clear agreement with the
relation (\ref{hyperscaling}) with $\qq = d/d_c = 1/2\sigma$, and inconsistent with
the conventional expectation $\qq = 1$.

Our simulation results for the finite-size scaling of the magnetisation and
susceptibility are depicted in Figs.~\ref{figm} and {\ref{figchi}}, respectively.
For the magnetisation, we fit the power-law \eqref{mL} to the data at the critical
point and for $L\ge 2^{14}$. We thus arrive at estimates
$-\qq\beta/\nu = -0.248 \pm 0.001$ for $\sigma = 0.1$ and $-\qq\beta/\nu = 
-0.249 \pm
0.001$ for $\sigma = 0.2$, respectively. Working at the maxima $T_L$ of the
susceptibility, on the other hand, we find $-\qq\beta/\nu = -0.248 \pm 0.001$ 
for
$\sigma = 0.1$ and $-\qq\beta/\nu = -0.246 \pm 0.001$ for $\sigma = 0.2$. With $\beta
= 1/2$ and $\nu = 1/\sigma$ from Eqs.~(\ref{abcd}) and (\ref{ef}), respectively,
these support $\qq = d/ d_c = 1/2\sigma$ over the alternative $\qq=1$.  For the
susceptibility, when $\sigma = 0.1$, we estimate the slope $\qq\gamma/\nu$ as $0.503
\pm 0.002$ at the critical point and $0.504 \pm 0.002$ at the pseudocritical point.
The equivalent results for $\sigma = 0.2$ are $0.503 \pm 0.002$ and $0.508 \pm
0.002$, respectively.  Again, $\gamma = 1$ and $\nu = 1/\sigma$ from
Eqs.~(\ref{abcd}) and (\ref{ef}), these support $\qq = d/ d_c = 1/2\sigma$.

\begin{figure}[t]
\begin{center}
\includegraphics[width=0.9\columnwidth, angle=0]{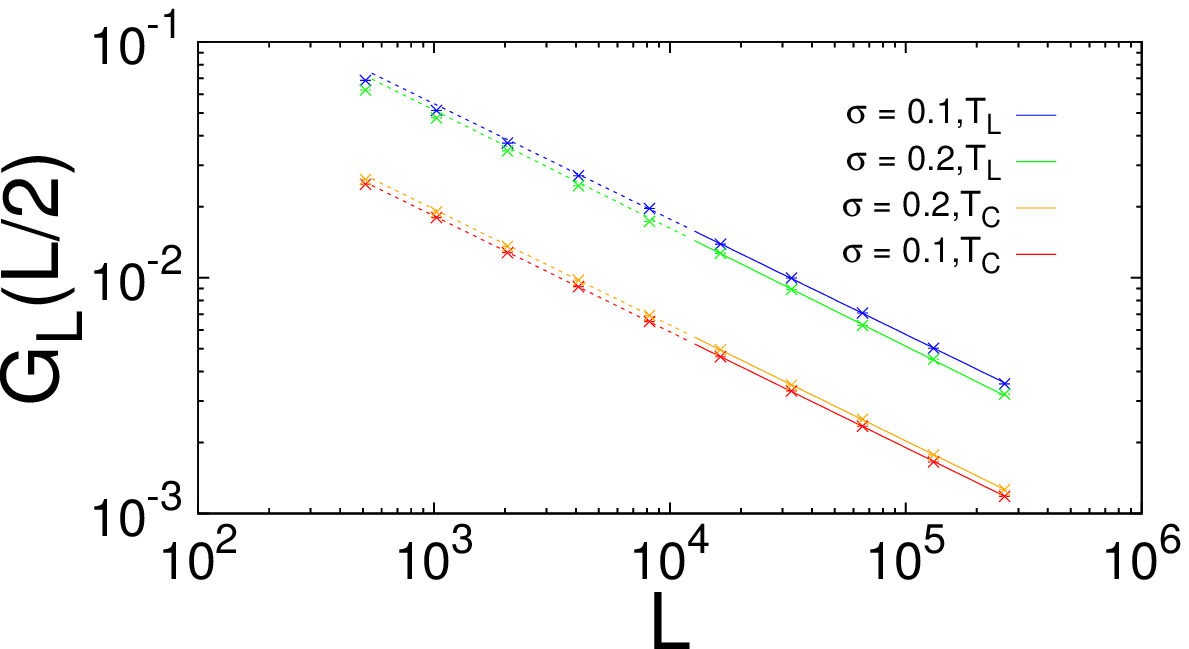}
\caption{ Finite-size scaling of the correlation function at distance $r=L/2$ for
  $\sigma = 0.1$ at the pseudocritical point (top set of data, blue online),  
$\sigma = 0.2$
  at the pseudocritical point (second from top, green online), $\sigma = 0.2$ 
at the
  critical point (third from top, orange online), $\sigma = 0.1$ at the
  critical point (lowest set of data, red online).  The slopes estimate
  $-d/2$ as $-0.496 \pm 0.001$, $-0.491 \pm 0.004$, $-0.490 \pm 0.002$ and 
$-0.498
  \pm 0.003$, respectively.  }\label{figG}
\end{center}
\end{figure}

Finally we address the decay of the correlation function at the critical and
pseudocritical points.  In Fig.~\ref{figG}, the correlation function at $r=L/2$ is
plotted against $L$ at both temperatures for $\sigma = 0.1$ and $\sigma = 0.2$.  The
conventional expectation comes from Eq.~(\ref{G1}) and gives $G_L(L/2) \sim
L^{-(d-2+\eta)} = L^{-(d-\sigma)} = L^{\sigma-1}$ in our case ($d=1$).  The $Q$-FSS
prediction comes from Eq.~(\ref{G11}) and is $ G_L(L/2) \sim L^{-(d-2+\eta_Q)} =
L^{-d/2} = L^{-1/2}$, independent of $\sigma$.  The measured exponents, again
restricting fits to the range $L\ge 2^{14}$, are $-0.498 \pm 0.003$ at the critical
point and $-0.496 \pm 0.001$ at the pseudocritical point for $\sigma=0.1$. The
corresponding measurements for $\sigma=0.2$ are $-0.490 \pm 0.002$ and $-0.491 \pm
0.004$, respectively.  Thus the $Q$-FSS prediction is supported.

\section{Discussion}
\setcounter{equation}{0}

The behaviour of the Ising model with long-range interactions remains a focus of
investigation in the study of fundamental properties of critical phenomena.  Here we
examined the model in circumstances where the interaction range is sufficiently long
for the model, although defined on a periodic chain, to be above its upper critical
dimension.  We have performed an extensive finite-size scaling study at both the
infinite-volume critical points and the finite-volume pseudocritical points and
confirmed that global quantities such as the magnetization and the susceptibility
scale with a modified version of finite-size scaling.  The origin of this
modification to standard FSS is the occurrence of dangerous irrelevant variables
above $d_c$.  Recent theoretical developments indicate that, contrary to
long-standing belief, dangerous irrelevant variables also alter the behaviour of the
correlation length and correlation function in high dimensions
\cite{BeKe12,ourCMP,ourEPL}. Here, we use extensive cluster-update Monte Carlo
simulations of the model close to criticality to investigate finite-size scaling in
this model and find that indeed this alteration does occur for interaction ranges
corresponding to the system being above its upper critical dimension.  In particular,
the algebraic scaling law $\xi_L \sim L^{\qqq}$ is supported, wherein $\qq = d/d_c$.
Turning to the decay of the correlation function on a finite-size lattice, the FSS
theory presented in \cite{ourEPL} shows that, above $d_c$, this is not captured by
the anomalous dimension $\eta$ derived from Landau or mean field theory.  We have
demonstrated this to be the case also in the one-dimensional model with sufficiently
long range and instead it is described by a new exponent $\eta_Q$.

~ \\ 
~ \\
\noindent
{\bf{Acknowledgements:}}

This research was supported by the Leipzig-Lorraine-Lviv-Coventry 
Doctoral College of the German-French University for the 
Statistical Physics of Complex Systems and by a 
Marie Curie (RAVEN, MC-IIF-300206) 
and IRSES grants 
            (SPIDER, PIRSES-GA-2011-295302 and DIONICOS, PIRSES-GA-2013-612707) within the 7th EU
Framework Programme. M.W.\
acknowledges funding in the Emmy Noether programme of the DFG under contract No.\
WE-4425/1.

\bigskip
%

\end{document}